\documentclass[journal]{IEEEtran}
\pdfoutput=1
\ifCLASSINFOpdf
\else
\fi
\usepackage{graphicx}

\usepackage{amssymb}

\usepackage{mathtools}

\usepackage{lineno}
\usepackage{xcolor}
\usepackage{amsthm}
\usepackage[nointegrals]{wasysym}
\usepackage{bbm}
\usepackage{array}
\usepackage{amsmath}
\usepackage{subfig}
\usepackage{soul,color}
\usepackage[ruled]{algorithm2e}
\usepackage{hyperref}

\newtheorem{definition}{Definition}
\theoremstyle{definition}
\newtheorem{proposition}{\textrm{Proposition}}

\newtheorem{assumption}{Assumption}

\newcommand{\argmax}{\mbox{argmax}}
\usepackage[noadjust]{cite}

\usepackage{titlesec}

\titleformat*{\section}{\large\bf}
\titleformat*{\subsection}{\normalsize\bf}
\titleformat*{\subsubsection}{\normalsize}

\titlespacing*{\section}{0pt}{2ex}{1ex}
\titlespacing*{\subsection}{0pt}{1ex}{0ex}
\titlespacing*{\subsubsection}{0pt}{1ex}{0ex}
\titlespacing*{\paragraph}{0pt}{1ex}{1em}

\begin{document}
%
\title{Auction Design through Multi-Agent Learning in Peer-to-Peer Energy Trading}
%
%
%

\author{Zibo~Zhao,
        Chen~Feng,
        and~Andrew~L.~Lu
\thanks{Zibo Zhao is with Google, Mountain View, CA, USA, e-mail: zibozhao.purdue@gmail.com. This work is based on his graduate research at Purdue University.}
\thanks{Chen Feng is with School of Industrial Engineering, Purdue University, West Lafayette, IN, USA, email: feng219@purdue.edu.}
\thanks{Andrew L. Lu is with School of Industrial Engineering, Purdue University, West Lafayette, IN, USA, email: andrewliu@purdue.edu.}}

\maketitle

\begin{abstract}
Distributed energy resources (DERs), such as rooftop solar panels, are growing rapidly and are reshaping power systems. To promote DERs, feed-in-tariff (FIT) is usually adopted by utilities to pay DER owners certain fixed rates for supplying energy to the grid. An alternative to FIT is a market-based approach; that is,  
consumers and DER owners trade energy in an auction-based peer-to-peer (P2P) market, and the rates are determined based on supply and demand. However, the auction complexity and market participants' bounded rationality may invalidate many well-established theories on auction design and hinder market development. To address the challenges, 
we propose an automated bidding framework based on multi-agent, multi-armed bandit learning for repeated auctions, which aims to minimize each bidder's cumulative regret. Numerical results indicate convergence of such a  multi-agent learning game to a steady-state. Being particularly interested in auction designs, we have applied the framework to four different implementations of repeated double-side auctions to compare their market outcomes. While it is difficult to pick a clear winner, $k$-double auction (a variant of uniform pricing auction) and McAfee auction (a variant of Vickrey double-auction) appear to perform well in general, with their respective strengths and weaknesses. 
\end{abstract}


%
\IEEEpeerreviewmaketitle

\section{Introduction}
\label{sec:intro}
Distributed energy resources (DERs) are a vital part of a smart grid, as such resources can improve system reliability and resilience with their proximity to load, and promote sustainability, as the majority of DERs being renewable energy resources \cite{jiayi2008review, akorede2010distributed}. To incentivize investments in DERs, two general approaches exist: non-market-based versus market-based. 
The most common and widely used policy in a non-market-based approach is feed-in-tariff (FIT) (including net-metering) \cite{lesser2008design}. While effective in promoting DERs, it may create equity issues as consumers without DERs would face increased electricity rates to pay for the FIT. In a market-based approach, DERs can choose to participate a wholesale electricity market, as specified in the recent FERC Order 2222 \cite{FERC2222}. To do so, however, an aggregator is needed to pool DER resources and to bid into a wholesale market on behalf of DER owners, as energy output from an individual owner (such as a household) is too small to bid into a wholesale market, nor do the owners have the expertise to do so. In addition, sending electricity from multiple and widely dispersed DER locations over long distance to a bulk transmission system will incur significant energy losses; nor does it contribute to the resilience of local distribution networks. An alternative market-based approach is to have a local marketplace (e.g., within the same distribution network) for consumers and DER owners, also referred to as prosumers, to directly trade energy, hence the so-called peer-to-peer (P2P) market. The actual rates that market participants pay/receive will fluctuate over time, reflecting the dynamic supply and demand conditions. 

In a bilateral marketplace, a leading mechanism to match supply and demand is through a double-side auction.  
While auction designs have been well studied in the field of economics and game theory \cite{friedman2018double, huang2002design, niu2013maximizing, nicolaisen2001market}, several special features of a peer-to-peer (P2P) energy market require special attention. First and foremost,  
a P2P energy market inherently involves repeated auctions and exogenous uncertainties (e.g., wind/solar availability), making the analysis of market participants' bidding/asking strategies much more difficult. Second, market participants are likely to have bounded rationality, such as no information of other players in the game. In fact, they may not even know their own valuation of energy production and consumption. This is especially so since the majority of DERs are wind or solar resources, which have significant investment costs but zero marginal costs. It is not clear how such resources should bid in a double auction. Furthermore, the participants' valuations are likely dependent, such as in a hot summer day, all buyers would value high of energy consumption for air conditioning. This feature alone would nullify the assumptions of most of the known results in auction theory, which require independent valuation among bidders \cite{MCMC87}.

There have been increasing amount of works to study double auctions in a P2P energy market setting, as surveyed in the review papers  \cite{PoorP2PReview,TransactiveEnergy_Review}. Here we focus on the works that are directly related to double auctions. In \cite{fuller2011_DA}, a double auction is proposed for residential users, with the focus on HVAC bidding. It uses a pre-determined demand curve (based on desirable temperature) to determine price and quantity bids for buyers; while for sellers, they all just bid a flat curve at some prevailing market price. This model is further extended in \cite{TCL_PartI,TCL_PartII}, which propose a  mechanism to implement a social choice function. However, these works focus on demand bids only, and the theoretical results require the knowledge of consumers' utility function. \cite{khorasany2017auction} considers explicitly a double-auction, but only uses an averaging mechanism to determine clearing prices -- adding all buyers and sellers bid/ask prices, then divided by the total number of bidders. Such a mechanism would make every bidder possible to manipulate clearing prices. 
\cite{StrategyProofMUDA} implements a variant of the McAfee auction, which will also be discussed in this paper,and proposes an approximate dynamic programming (ADP) approach to help bidders (mainly with battery resources) bid. The reward function in the ADP is set to be the economic cost of prosumers/energy storage owners, which can be ambiguous for zero-marginal cost resources. This is the view shared both by us and in  \cite{CompareAuctions}, which is closest to our works. \cite{CompareAuctions} recognizes several issues that are also emphasized in this paper, including the zero marginal cost issue, and numerically compares three auction mechanisms: uniform-pricing, Vickrey and pay-as-bid, all of which will be analyzed in this paper. However, \cite{CompareAuctions} only let sellers (with zero-marginal-cost resources) bid levelized energy costs, which are just levelized investment costs of the resources. This neither is a strategy nor does it have theoretical justifications, as sunk costs should not be factored into operation/bidding decisions. 

Despite the existing works, significant knowledge gaps remain before actual implementation of a P2P market. Most importantly, no studies have focused on the repeated nature of local energy trading, and the potential of market participants (or machines/algorithms) to learn to overcome their bounded rationality and to adapt to the unique features such as all zero-marginal-cost supplies, where the well-established marginal pricing framework in economics would not work. To fill-in the gaps,  and to provide an algorithmic-framework that can achieve control automation to aid consumers/prosumers bid into a repeated double-auction market, we propose a multi-agent, multi-armed bandit (MAB) learning approach (herein referred to as MAB games).\footnote{In this paper, we refer to all market participants, including pure buyers, pure sellers, and prosumers, who may buy or sell energy at a given moment, as agents.} 
The essence of an MAB game is that each agent uses an algorithm to choose an action (referred to as an arm) from a finite number of available actions at each round of decision-making (a round of auction in our context). The collective actions of the agents determine the reward for each agent, who can compare it with the best-possible reward they could get in hindsight, with the difference being called regret. The agents then seek to find an algorithm to minimize cumulative regrets. While regret-minimizing algorithms have been well-established for single-agent MAB problems,
such as the famed Upper Confidence Bound (UCB) algorithm \cite{auer2002finite}, multi-agent MAB games have only received attention recently, and is a special case of a multi-agent reinforcement learning game, as surveyed in \cite{BasarMARL}. 
A fundamental difficulty of any multi-agent learning-based approach is the lack of stationarity of the system (such as market prices) due to the multi-agent interaction; while stationarity is a key assumption for most of the regret-minimizing MAB algorithms to converge. In \cite{gummadi2016mean}, a mean-field approach is proposed to address the theoretical issue of an MAB game, which contains a large number of agents, and each agent believes that the system is stabilized at a mean field steady state (MFSS) nd their individual actions will not affect the steady state. Under this belief, \cite{gummadi2016mean} shows the existence and uniqueness of the MFSS, and the convergence to MFSS in a multi-agent MAB game. While we cannot establish such results in our work due to technical difficulties to be discussed in the paper, our numerical results nevertheless indicate convergence in all the instances. 

In this paper, we demonstrate the flexibility of  the MAB-game approach in modeling multi-agent repeated decision-making and use it to compare four specific double-auction mechanisms: uniform-price Vikrey, McAfee, and maximum volume matching (pay-as-bid). Independent of the MAB game framework, we establish theoretical properties of the four auctions with the unique features presented by a local P2P energy market, including the public knowledge of both buyers and sellers reservation price (i.e., the rates to buy from or sell to a utility company if not traded in the P2P market) and zero-marginal costs of all suppliers. Our theoretical results show that the unique features can bring both desirable and undesirable auction outcomes, which are attested by simulation results from the MAB-game approach.  

While the models in this paper do not have physical modeling of an electric distribution network, we want to point out that it is not the limitation of the MAB-game framework. In another line of our work \cite{zhao2018electricity}, we used the MAB-game approach to model consumers' decentralized response to real-time prices of wholesale energy markets (aka the locational marginal prices or LMPs). There we explicitly embedded a system operator's optimal power flow problem into the MAB-game, and showed (numerically) that agents can even learn to alleviate congestion at different time and locations to reduce their electric bills. These being said, incorporating distribution network constraints in a P2P double auction is not a trivial matter at all. We will briefly discuss this in the paper and point that as a future research direction. By the same token, we do not discuss specifically the role a utility company or a distributed system operator (DSO) would play in a P2P market. In our view, their roles will be more on maintaining system reliability than on the financial transactions among market participants. 

The rest of the paper is structured as follows. Section \ref{sec:mab-game} lays out the details of the MAB-game framework to study repeated double-auctions in a P2P energy market. Section \ref{sec:double_auction} introduces four specific double-auction mechanisms, and their theoretical properties (in the specific setting of a local energy market) are discussed in Section \ref{sec:analysis}. Numerical results are presented in Section \ref{sec:num_res}; while limitations of the current work and possible future research are discussed in Section \ref{sec:conclusion}.

\section{Learning in Double Auctions}
\label{sec:mab-game}
Consider a local electricity distribution network. Without a marketplace, prosumers can only sell their generated energy to the utility or a DSO at some pre-defined fixed FIT, denoted as $P_{FIT}$. Similarly, consumers can only buy energy from the utility at the utility rate (UR), denoted as $P_{UR}$. Throughout this paper, it is assumed that $P_{FIT} < P_{UR}$.\footnote{The current FIT rates by states can be found at \url{https://www.eia.gov/electricity/policies/provider_programs.php}. While the FIT in some states is lower than UR, it is higher in some other states. In the later case, we argue that it is not a fair policy as the DER owners are over compensated at the expense of the rest of the consumers (assuming that the utilities will pass the FIT to all consumers).} With an auction-based marketplace, energy buyers would like to pay prices lower than $P_{UR}$; similarly, energy sellers want to receive better rates than $P_{FIT}$. 

\subsection{Agents and Types}\label{subsec:AgentType} We consider three kinds of market participants: pure buyers, pure sellers, and prosumers. For the last group, there role is not fixed; namely, an prosumer can be either a buyer or a seller at any particular round of an auction, just not at the same time. More specifically, let $\mathcal{A}^h_b$ denote the set of buyers at round $h$ of an auction (such as at a particular hour $h$ in a particular day), and $\mathcal{A}^h_s$ be the set of sellers at the same round. Then $\mathcal{A}^h_b \cap \mathcal{A}^h_s = \emptyset$. Furthermore, let $\mathcal{A}^h = \mathcal{A}^h_b \cup \mathcal{A}^h_s$. Not only the sets $\mathcal{A}^h_b$ and $\mathcal{A}^h_s$ may change over $h$, so is the joint set $\mathcal{A}^h$. In addition to account for prosuemers' altering positions, the changing agent sets also reflect situations where some agents may leave the auctions (such as moving out of the local network) and new agents may join. We want to highlight this capability and flexibility of dealing with dynamic agents as a particular benefit of the learning-based approach. 

In a double-side auction, both buyers and sellers need to decide their bid/ask prices and quantities of energy. As a starting point, we do not consider storage options in this work (and will discuss the challenges of considering storage and potential solutions in the conclusion section). Without storage, the energy quantities produced by DERs are likely not controllable, and hence we assume that any quantities not needed by the DER owner (aka the prosumer) will be sold to the local market.\footnote{This is exactly the case of a  grid-tied solar system, where the grid essentially serves as a battery.} Consequently, in our auction setting, agents only bid/ask (per unit) energy prices. While the agents do not control quantities; their demand and output quantities from DERs are still stochastic, reflecting the generation variations from renewable resources. More modeling details are introduced in the following subsections. We use a generic variable $y^h_i$ to denote the type of agent $i\in \mathcal{A}^h$ in a particular round $h$, which specifies what kind of market participant $i$ is, as well as the distributions of their energy consumption and generation quantity (if $i$ is a pure seller or prosuemr) at $h$. The set of all possible types for an agent $i$ is denoted by $Y_i$.

\subsection{Discrete Price Arms}
\label{subsec:price_arms}
Since the majority of DERs are solar and wind resources, we assume that the sellers' marginal costs are all 0.\footnote{Note that we actually do not need any assumptions on marginal costs, as any rational sellers will bid at or above their marginal costs, and this would not affect our modeling in anyway. Assuming zero marginal costs just makes it easier to gain insights from the numerical results.} Hence, any rate higher than FIT would be attractive to DER owners. Similarly, energy buyers would desire for any rate lower than the UR rate. In another words, any rate in the range of $(P_{FIT}, P_{UR})$ would be preferred by both the buyers and sellers. To implement a learning-based algorithm, we discretize the interval $[P_{FIT}, P_{UR}]$ into $M$ elements, such as by quarter or dollar increments, and refer to each element $p(m)\in [P_{FIT}, P_{UR}]$, $m =1 , \ldots, M,$ a price arm. At each round of an auction, each buyer/seller chooses a price arm to bid/ask. Since the zero marginal-cost of energy production is common-knowledge to all agents, as well as the de facto price ceiling ($P_{UR}$) and price floor ($P_{FIT}$), the agents (both buyers and sellers) try to learn how to choose the price arms in the repeated auction to maximize their rewards, with the rewards being explicitly defined in the next subsection, and the learning algorithm introduced in Subsection \ref{subsec:pricing_bandit_lrn}. 

\subsection{Rewards}
\label{subsec:rwd_normal}
Conceptually, the (marginal) rewards for buyers in one-round of the auction is the difference between the prices they pay for one unit of energy and $P_{UR}$; similarly, the (marginal) rewards for sellers in one round is the difference between the prices they are paid and $P_{FIT}$. To aid the model development, it is more convenient to scale the agents' rewards between 0 and 1. To do so, we first define two benchmarking payoffs; that is, the lower and upper bound of an agent's payoff. (Note that the concept of payoffs for sellers are unambiguous; for buyers, payoffs should be understood as payments to purchase energy. They are negative numbers in our modeling setup, to be explained next; and hence, it is still that the bigger the payoff (meaning closer to 0), the better for a buyer.) 

Let $q^h_i$ denote agent $i$'s actual demand or generation (not just bid/ask quantity) at a round $h$ of an auction. To ease the arguments, we drop the round (or time) superscript in the notation within this subsection, and it is understood that all discussions are within one round of the repeated auction. The quantity $q_i$ is negative for a buyer and positive for a seller; namely, $q_i > 0$ for ${i \in \mathcal{A}_b}$ and $q_i < 0$ for ${i \in \mathcal{A}_s}$. 
For a  buyer agent, the lower bound of the payoff is naturally to pay all of $q_i$ at $P_{UR}$; for the upper bound, we define it to be $q_i \cdot P_{FIT}$, as no sellers would be willing to supply energy at a rate lower than $P_{FIT}$. (Note that $q_i$ is negative for a buyer;  hence $q_i P_{FIT} > q_i P_{UR}$ with the assumption that $P_{FIT} < P_{UR}$). For sellers, the lower and upper bounds are exactly reversed. To avoid discussing the buyers and sellers separately, we can use the indicator function $\mathbbm{1}_{\{ \}}$ to define a uniform set of notations. Specifically, we can define the lower and upper bound of payoff for any agent $i \in \mathcal{A}$ as follows:
\begin{equation}
	\label{eq:rwd_low}
	\underline{\Lambda_i} = q_i \cdot [P_{UR} \cdot \mathbbm{1}_{\{ i \in \mathcal{A}_b\}} + P_{FIT} \cdot \mathbbm{1}_{\{ i \in \mathcal{A}_s\}}],
\end{equation}
\begin{equation}
	\label{eq:rwd_high}
	\overline{\Lambda_i} = q_i \cdot [P_{FIT} \cdot \mathbbm{1}_{\{ i \in \mathcal{A}_b\}} + P_{UR} \cdot \mathbbm{1}_{\{ i \in \mathcal{A}_s\}}],
\end{equation}
where $\mathbbm{1}_{\{ i \in \mathcal{A}_b\}} = 1$ if agent $i$ is a buyer (i.e., $i\in \mathcal{A}_b$), and it equals 0 otherwise. The definition for $\mathbbm{1}_{\{ i \in \mathcal{A}_s\}}$ is the same.

Now let $\Lambda_i$ denote the actual payoff of agent $i$ in a round of the auction, which should satisfy that $\underline{\Lambda_i} \leq \Lambda_i \leq 	\overline{\Lambda_i}$. Throughout this paper, we assume that for the uncleared bid or ask quantities in an auction, they will be purchased by a utility company at $P_{UR}$ or sold to a utility at $P_{FIT}$. Note that depending on the specific auction design, there can be partial clearing, meaning that for the same bidder, only part of their  bid quantity may be cleared. Hence, within any round, an agent's payoff consists of two components: the payoff from participating the auction (denoted as$\Lambda_i^{au}$), and the payoff from selling to or buying from the utility (denoted $\Lambda_i^{ut}$); that is, 
\begin{equation}
	\label{eq:rwd_origin}
	\Lambda_i = \Lambda_i^{au} + \Lambda_i^{ut}. 
\end{equation} 
More specifically, let $q_i^{au}$ denote agent $i$'s cleared quantity in an auction ($q_i^{au} \geq 0$ for $i\in\mathcal{A}_s$, and $q_i^{au} \leq 0$ for $i\in\mathcal{A}_b$), and $p_i^{au}$ be the corresponding unit price determined by an auction for agent $i$ to receive (if a seller) or to pay (if a buyer). 
Then $\Lambda_i^{au} = p_i^{au}\cdot q_i^{au}.$ Similarly, we have that $\Lambda_i^{ut} = p_i^{ut}\cdot q_i^{ut}$, where $p_i^{ut} = P_{FIT}$ if $i \in \mathcal{A}_s$, and $p_i^{ut} = P_{UR}$ if $i \in \mathcal{A}_b$, and $q_i^{ut}$ denotes the uncleared energy quantity for agent $i$.

With the above notations, we can now define the normalized reward as follows, 
\begin{equation}
\label{eq:rwd_normal}
\pi_i = ( \Lambda_i - \underline{\Lambda_i} ) / ( \overline{\Lambda_i} - \underline{\Lambda_i} ).
\end{equation}
It is straightforward to see that if agent $i$'s auction clearing price $p_i^{au}$ is in $[P_{FIT}, P_{UR}]$, we have $\Lambda_i \in [\underline{\Lambda_i}, \overline{\Lambda_i}]$ and hence $\pi_i \in [0, 1]$. If agent $i$ is not cleared in an auction at all, then $\Lambda_i =  \underline{\Lambda_i}$ and $\pi_i = 0$. 

Note that in our simulation setup, we do allow sellers to ask above $P_{UR}$, and buyers to bid below $P_{FIT}$. This is to represent the case that either the auction agents have bounded rationality (in the sense that they do not recognize the practical price ceilings/floors) or the agents are greedy, as if they want to try their luck to earn a higher payoff in one round of the auction. Our numerical results later show that indeed the auction clearing price could go beyond $P_{UR}$ or below $P_{FIT}$ in certain rounds. (But such clearing prices cannot be sustained in the repeated auctions as the counterpart agents can quickly learn to reverse the course.) To prevent the normalized reward to go outside the range of $[0, 1]$, we expand the definition of $\pi_i$ for all $i \in \mathcal{A}$ as follows: 
\begin{equation}
\label{eq:rwd_function}
\pi_i=
\begin{cases}
1 \cdot \mathbbm{1}_{\{ i \in \mathcal{A}_b\}} + 0 \cdot \mathbbm{1}_{\{ i \in \mathcal{A}_s\}}, & \text{for } p_i^{au} < P_{FIT}\\
( \Lambda_i - \underline{\Lambda_i} ) / ( \overline{\Lambda_i} - \underline{\Lambda_i} ), & \text{for }  P_{FIT} \le p_i^{au} \le P_{UR}\\
0 \cdot \mathbbm{1}_{\{ i \in \mathcal{A}_b\}} + 1 \cdot \mathbbm{1}_{\{ i \in \mathcal{A}_s\}}, & \text{for } p_i^{au} > P_{UR}
\end{cases}.
\end{equation}
With \eqref{eq:rwd_function}, $\pi_i$ will always be within $[0, 1]$, regardless what the clearing price is for agent $i$, or if agent $i$ is cleared at all. We will use the reward defined in \eqref{eq:rwd_function} to set up the learning-based framework below. 

\subsection{Policies and Regret Minimization}
\label{subsec:pricing_bandit_lrn}

Once the reward of each agent is defined, each agent learns from the history of the game to decide what to do in the next round. The history of the game for agent $i$ is recorded in the state variables. For agent $i$ at the $h$-th round of the repeated auction, the state variable, denoted by $z^{h}_i$, is a vector of $2M$ elements, with $M$ defined earlier as the number of price arms. The first $M$ elements record the number of times that each arm $m\in \{1, \ldots, M\}$ has been chosen by agent $i$; while the second $M$ elements  denote the average rewards (from $h=1$ to the current round $h$) associated with each arm $m$.  
Let $\mathcal{Z}^{h}_i$ denote the set of all possible states for agent $i$ at round $H$. 
A policy, also referred to as a strategy or simply an algorithm, is a mapping from $\mathcal{Z}^{h}_i$ to a probability distribution over the arms. Specifically, let $\Xi =\{\xi=(\xi_1, \ldots,  \xi_M): \sum_{m=1}^M \xi_m =1\}\in[0,1]^M $ be the set of probability distributions over the $M$ price arms to choose from for each agent $i$. Then agent $i$'s policy, denoted by 
$\sigma_{i}$, is a mapping from $\mathcal{Z}^{h}_i$ to $\Xi$.  For agents employing a policy as defined above, the actual arm that a consumer $i$ will choose in round $h$ is then a random variable. One example of a policy is the $\epsilon$-greedy algorithm -- at each round $h$, an agent $i$ chooses the arm $m$ with the highest average reward so far with probability $1-\epsilon$, and randomly chooses another arm with probability $\epsilon$, with $\epsilon \in (0, 1)$. 

Note that an agent's reward at each round of the auction does not only depend on which arm they choose, but also depends on the collective actions of all agents. The outcomes of all agents' actions in a specific round $h$ can be represented by a quantity referred to as the  population profile, which is the histogram of the arm choices by all the agent at round $h$. Let $\mathbf{f}^{|\mathcal{A}^h|}(m)$ denote the population profile for a specific arm $m$, where $|\mathcal{A}^h|$ denotes the number of total agents at round $h$. Then 
\begin{equation}\label{eq:PopProfile} 
 \mathbf{f}^{|\mathcal{A}^h|}(m) = \frac{1}{|\mathcal{A}^h|}\sum_{i=1}^{|\mathcal{A}^h|} \mathbbm{1}\{\sigma_i(z_i^h)= m\},
\end{equation}
where the function $\mathbbm{1}\{\sigma_i(z_i^h)= m\} = 1$ if agent $i$ chooses the arm $m$ at round $h$, and is zero otherwise.

A key technical difficulty with a multi-agent MAB game is that the population profile is both random and does not follow a stationary distribution. The essence of a mean-field approach is to assume that each agent believes that the population profile is in a steady state, denoted as  $\boldsymbol{f} = \{f(m)\}_{m=1}^M$. Under a stationary population profile, we can define the best possible reward in one round for agent $i$ as follows: 
\begin{equation}\label{eq:BestReward}
	\pi_i^*(\boldsymbol{f}) 
	= 
	\max_{m = 1, \ldots, M} \mathbbm{E} [\pi_i( \boldsymbol{f}, m)], 
\end{equation}
where $\pi_i(\boldsymbol{f}, m)$ denotes the reward of agent $i$ for choosing price arm $m$ under the population profile $\boldsymbol{f}$. 

Let $\sigma_i$ be an arbitrary policy for agent $i$ to choose an arm at each round. Over $D$ rounds, we use $\Gamma_{\sigma_i}(D, m)$ to denote the number of times that price arm $m$ has been chosen by the policy $\sigma_i$. Then we can define agent $i$'s cumulative regret under the policy $\sigma_i$ as below:
\begin{equation}
	\label{eq:cum_regret}
	\Delta_{\sigma_i} 
	= 
	\pi_i^*(\boldsymbol{f}) \cdot D 
	- 
	\sum_{m = 1, \ldots, M} \mathbbm{E} [\pi_i(\boldsymbol{f}, m) \cdot \Gamma_{\sigma_i}(D, m)].
\end{equation}

 
The regret $\Delta_{\sigma_i}$ in Eq. (\ref{eq:cum_regret}) is the expected cumulative loss due to the fact that the policy $\sigma_i$ does not necessarily always pick up the optimal price arm under the stationary population profile $\mathbf{f}$, which is unknown to the agent. A policy $\sigma_i$ is called a \textit{no-regret} policy if the regret in  Eq. (\ref{eq:cum_regret}) satisfies:
\begin{equation}
\frac{1}{D}\Delta_{\sigma}  < R(D, M),
\end{equation}
where the function $R$ is $o(1)$ with respect to $D$, and $M$ is the total number of arms. Regret-minimizing policies for a single-agent MAB problem have been well-studied (under the assumption that the distribution of the exogenous uncertainty is stationary). One popular policy is the so-called UCB (Upper Confidence Bound) algorithm \cite{auer2002finite}, whose idea is simple: at the $D$-th round of the game, an agent chooses the arm $\hat{m}$, with 
\begin{equation}\label{eq:UCB}
	\hat{m} \in \displaystyle \argmax_{m\in\{1, \ldots, M\}} \left\{\overline{\Pi}_i(m) + \sqrt{\frac{2\ln(D)}{\Gamma_i(D, m)}}\right\},
\end{equation}
where $\overline{\Pi}_i(m) $ represents the average reward for agent $i$ when the arm $m$ is chosen up to round $D$, and $\Gamma_i(D, m)$ again is the total number that arm $m$ has been chosen up to $D$. The objective function in \eqref{eq:UCB} reflects the trade-off between exploitation (choosing an arm with the largest  payoff) and exploration (trying as many arms as possible).

In \cite{gummadi2016mean}, it has been proven that under the assumption that each agent's reward is continuous with respect to the population profile, then a mean-field steady state (MFSS) exists for any policy. In addition, they show that if the reward function is also Lipschitz continuous with respect to the population profile, then the MFSS is unique, and as the number of agents grows to infinity, the dynamics of the multi-agent system will converge to the unique MFSS. These results are really strong and desirable, as it shows the robustness of the approach (with a unique MFSS to converge to) and its scalability, which overcomes the computational issues of pure agent-based simulation that may not scale well with more agents. However, the key continuity assumption does not hold in the auction situation, as each agent's reward depends on if they win or lose the auction, which can change abruptly with a small change in population profile (i.e., how other agents bid). Despite the lack of theoretical results, our numerical experiments do show the emergence of a steady-state population profile in all the simulations.

\section{Double Auction Designs}
\label{sec:double_auction}
In this section, we lay out four specific auction designs to realize a multi-unit, double-side auction; they are are the $k$-auction, a variant of Vickrey double-side auction \cite{huang2002design}, McAfee auction \cite{mcafee1992dominant} and the maximum volume matching auction \cite{niu2013maximizing}. The auction designs cover the cases of both uniform pricing (i.e., all cleared sellers receive the same price, and so do all the cleared buyers) and differential pricing (i.e., cleared buyers and sellers all receive different prices). While the auction design issue was hotly debated at the beginning of wholesale electricity markets \cite{ElecAuction}, with uniform pricing emerging triumphantly, we do want to study if uniform-pricing-based auction design still works in the case when all the sellers' marginal costs are zero. 

\subsection{$k$-Double Auction}\label{subsec:k-double}
To start with, all bids/asks are sorted by their bidding/asking prices, which result in the stair-wise demand/supply curves as shown in Figure \ref{fig:kdouble}. At the quantity of the critical intersection point $Q^*$ where the aggregate demand and supply meet, the last step of the cleared bids from buyers are denoted as $(pb_{L}, qb_{L})$, and highest cleared asks from sellers are denoted as $(ps_{H}, qs_{H})$. The $k$-double auction represents a whole class of auctions with similar designs, with the parameter $k$ ranging from 0 to 1 in setting the market clearing price. More specifically, let $P^*$ denote the market clearing price in a $k$-double auction, then 
\begin{equation}\label{eq:k-Price} 
	P^* = k pb_L + (1-k) ps_H.
\end{equation} 
\begin{figure}
	\begin{center}
		\subfloat{\includegraphics[width=3.5in]{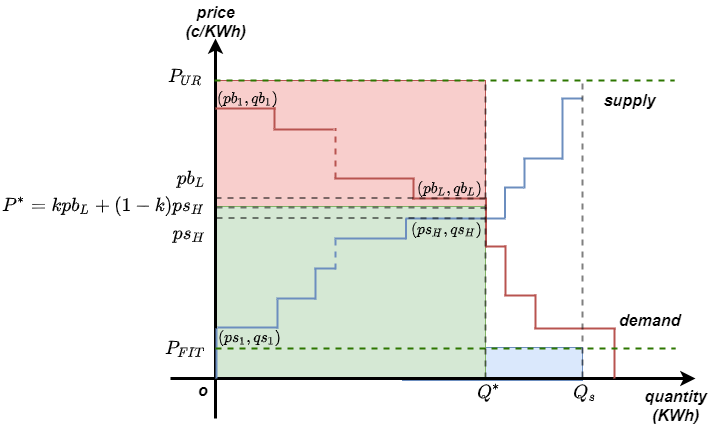}%
			\label{fig:kdouble}}\\
	\end{center}
	\caption{$k$-double auction.}
	\label{fig:kdouble}
\end{figure}
For the clearing mechanism to work under any supply/demand conditions, we define two rules as follow.
\subsubsection*{\textbf{Rule 1}}
If $\sum_{l=1}^{L} qb_l \ge \sum_{h=1}^{H} qs_h$ (referred to as over-demand), all the asks with $h \leq H$ are cleared, and sell all their quantities at price $P^*$; all the bids with $l \leq L$ are cleared, and the clearing price for all the buyers is also $P^*$. However, the quantity for each cleared buyer $l$ is not the quantity they bid; it is scaled back by the amount of over-demand. Specifically, for each buyer $l \leq L$, their cleared quantity is $qb_l - (\sum_{l=1}^{L} qb_l - \sum_{h=1}^{H} qs_h) / L$. As mentioned earlier, uncleared supply is assumed to sell to a utility at $P_{FIT}$; while uncleared demand is to buy at from the utility at $P_{UR}$. This is the same across all the rules and all the auctions below, and hence will not be stated again.

\subsubsection*{\textbf{Rule 2}}
If $\sum_{l=1}^{L} qb_l \le \sum_{h=1}^{H} qs_h$ (referred to as over-supply), all the buyers' bids with $l \leq L$ are cleared and buy all their demand bids  $qb_l$ at the price $P^*$; all the supply asks with $h \leq H$ are cleared and sell at the price $P^*$, but their cleared quantities are scaled back to $qs_h - (\sum_{h=1}^{H} qs_h - \sum_{l=1}^{L} qb_l)/ H$. 


To aid the comparison among different auction designs, we define the concept of auction surplus as the total surpluses of buyers and sellers, and denote it as $\widehat{S}$. We also denote the auctioneer's surplus by $S_{au}$. 
In a $k$-double action, since the market clearing price $P^*$ is the same for both buyers and sellers, clearly the auctioneer has zero profit; that is,  $S_{au}^{k-double} = 0$. 
For buyers, we define their surplus as $\sum_{i\in\mathcal{A}_{b}} (\Lambda_i - \underline{\Lambda}_i)$, with $\Lambda_i$ and $\underline{\Lambda}_i$ being defined in \eqref{eq:rwd_low} and \eqref{eq:rwd_origin}, respectively. 
The buyers surplus in a $k$-double auction is the upper rectangle (of dark pink color) in Figure \ref{fig:kdouble}. (If a buyer $i$ is not cleared in a $k$-double auction, $\Lambda_i = \underline{\Lambda}_i$ by definition, which means zero surplus.) We want to point out that the buyers' surplus here is different than the classic concept of consumer surplus, which refers to the difference between consumers' willingness-to-pay and the market clearing price (that is, the area below consumers' aggregated 
 demand function and above the market clearing price). The difference is important here because the buyers bid curve in Figure \ref{fig:kdouble} do not necessarily represent their true willingness-to-pay, nor do we make such assumptions. This is a particular benefit of the learning-based approach, from our perspectives, as we believe that the concept of willingness-to-pay (or consumers' utility function) is elusive and impractical in practice, since consumers likely do not know their true willingness-to-pay for electricity, which also changes over time. 
 
 For sellers, since we assume that all sellers' marginal costs are zero, their surpluses equal price times quantity, with the price being set by an auction if the supply ask is cleared, or being $P_{FIT}$ if the ask is not cleared. In Figure \ref{fig:kdouble}, suppliers surplus is the green area plus the blue area. Note that sellers surplus is different than the reward definition in \eqref{eq:rwd_function}. By \eqref{eq:rwd_function}, sellers' rewards will be 0 if their asks are not cleared in an auction. Hence they are motivated to learn to adjust their asks in the following rounds of the auction to aim to be cleared and to receive more a favorable price than $P_{FIT}$. 
 
 With the above definition of buyers and sellers' surplus, we can write their surplus in a simple way for the $k$-double auction as follows: 
\begin{equation}
\widehat{S}^{k-double} = P_{UR} \cdot Q^* + P_{FIT} \cdot (Q_s - Q^*).
\end{equation}

\subsection{Vickrey Variant Double Auction}\label{subsec:Vickrey}
Vickrey auction \cite{vickrey1961counterspeculation} is arguably one of the most famed auction design, mainly due to its truth-revealing property (which we will discuss in the next section). The original Vickrey auction has been extended to a double-side version for multiple-units goods in \cite{huang2002design}. Here we refer to it as the Vickerey-like auction, and introduce the basic set up of the auction here. More details can be found in the original paper \cite{huang2002design}. 
Similar to the $k$-double auction, all bids/asks are sorted by price, which result in the stair-wise demand/supply curves as shown in Figure  \ref{fig:vickrey_like}, 
\begin{figure}[t]
	\begin{center}
		\subfloat{\includegraphics[width=3.5in]{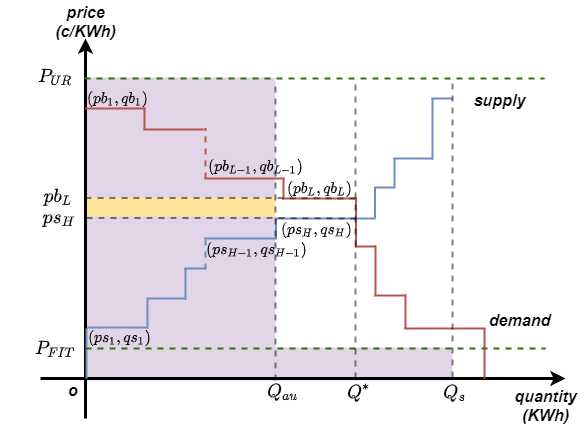}%
			\label{fig:vickrey_like}}\\
	\end{center}
	\caption{A Vickrey-like double auction market (Case I).}
	\label{fig:vickrey_like}
\end{figure}
At the critical intersection point $( P^*, Q^*)$ where the aggregate demand and supply meet, the last step of the cleared bids from buyers are denoted as $(pb_{L}, qb_{L})$, and highest cleared asks from sellers are denoted as $(ps_{H}, qs_{H})$. Now we consider two cases. Case I is exactly shown as  in Figure \ref{fig:vickrey_like}), where 
\begin{equation}\label{eq:CaseI}
pb_L \ge ps_H \ge pb_{L+1},\ \mathrm{and}\ \sum_{h = 1}^{H-1} qs_{h} 
\le
\sum_{l = 1}^{L} qb_{l}
\le
\sum_{h = 1}^{H} qs_{h}.
\end{equation}
Case II covers the other possibility at clearing, where
\begin{equation}\label{eq:CaseII}
ps_{H+1} \ge pb_L \ge ps_{H},\ \mathrm{and}\ 
\sum_{l = 1}^{L-1} qb_{l} 
\le
 \sum_{h = 1}^{H} qs_{h}
 \le
 \sum_{l = 1}^{L} qb_{l}.
\end{equation}

Here we describe the clearing mechanism for Case I only, as the mechanism for  Case II is the same. While the rules are similar to those in the $k$-double auction case, the key difference is that only up to $H-1$-th seller and $L-1$-th buyer are cleared, and their clearing prices can be different. 

\subsubsection*{\textbf{Rule 1}}
If $\sum_{l=1}^{L-1} qb_l \ge \sum_{h=1}^{H-1} qs_h$, all the asks with $h \leq H - 1$ are cleared, and sell all their supply asks at price $ps_H$; all the bids with $l \leq L-1$ are cleared, and the clearing price for all the buyers is $pb_L$; while the quantity for each cleared buyer is $qb_l - (\sum_{l=1}^{L-1} qb_l - \sum_{h=1}^{H-1} qs_h) / (L - 1)$. This feature of clearing only up to the $L-1$-th buyer and $H-1$-th seller is exactly inspired by the Vickrey auction to induce the truth revealing property (that is, no agents have the incentive to not bid truthfully).

\subsubsection*{\textbf{Rule 2}}
If $\sum_{l=1}^{L-1} qb_l \le \sum_{h=1}^{H-1} qs_h$, all the buyers bids with $l \leq L-1$ are cleared and buy all their demand bids $qb_l$ at price $pb_L$; all the supply asks with $h \leq H - 1$ are cleared and sell at $ps_H$; while the quantity to sell is $qs_h - (\sum_{h=1}^{H-1} qs_h - \sum_{l=1}^{L-1} qb_l)/ (H - 1)$. 

Based on the clearing rules, the total cleared quantity in the Vickrey-like auction (denoted by $Q^V$) is 
\begin{equation}
Q^V := min(\sum_{l=1}^{L-1} qb_l,\ \sum_{h=1}^{H-1} qs_h).
\end{equation}

Regarding to auction surplus, the auctioneer now has a positive surplus due to the difference between the cleared buying and selling price (as illustrated in the yellow shaded area in Figure \ref{fig:vickrey_like}). Mathematically,   
\begin{equation}
	S_{au}^{V}
	= (pb_L - ps_H) \cdot Q^V.
\end{equation}
The total agents' surplus in a Vickrey-like auction can be calculated as follows (represented by the light purple area in Figure \ref{fig:vickrey_like}):
\begin{equation}
\widehat{S}^{V}
=
[(P_{UR} - pb_L) + ps_H]\cdot Q^V
+
P_{FIT} \cdot (Q_s - Q^V).
\end{equation}

\subsection{McAfee's Double Auction}\label{subsec:McAfee's}
This mechanism is a variant of the Vickrey-like auction, suggested by
McAfee \cite{mcafee1992dominant}. Consider $P_0 = \frac{1}{2}(pb_{L+1} + ps_{H+1})$ and there are two cases: Cases A: if $P_0 \in [ ps_{H}, pb_{L}]$, the clearing mechanism works the same as the $k$-double auction, with a uniform clearing price of $P^* = P_0$ (shown in Fig. \ref{fig:McAfee}). The first $L$ buyers and first $H$ sellers are cleared.  
Case B: if $P_0 \notin [ ps_{H}, pb_{L}]$, the mechanism implements the Vickrey-like auction as in Section \ref{subsec:Vickrey} (shown in Fig. \ref{fig:vickrey_like}), and clears only up to the $L$-1-th buyer and $H$-1-th seller.

\begin{figure}[t]
	\begin{center}
		\subfloat{\includegraphics[width=3.5in]{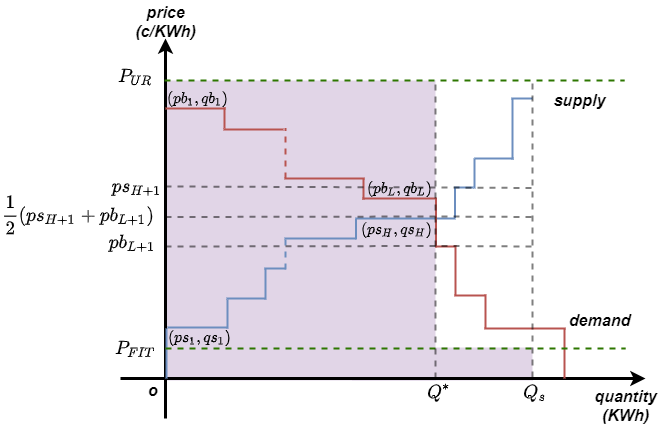}%
			\label{fig:McAfee}}\\
	\end{center}
	\caption{McAfee double auction market (Case A).}
	\label{fig:McAfee}
\end{figure}

\subsection{Maximum Volume Matching Double Auction}\label{subsec:MVM}
While the above three mechanisms differ in the details of how market clearing prices and quantities are determined, they all build upon the same idea of stacking supply asks and demand bids to find the intersection point, which originates from the classic economic idea of social welfare maximization. 
With a drastically different idea, \cite{niu2013maximizing} proposes a pay-as-bid auction (as opposed to a uniform-price auction) whose sole purpose is to maximize the cleared volume of the traded goods. 
Such an auction is referred to as the maximum volume matching (MVM) auction. 
The idea of maximizing traded volume is very appealing in our context, as it could help promote the penetration of renewable energy at the distribution level. 
Hence, we include the MVM mechanism in our comparison study. The mechanism of the market clearing process of the MVM auction is as follows. 
We start with the regular stacked supply asks/demand bids curves as in Figure \ref{fig:kdouble}. Then the supply stack is flipped horizontally around the vertical axis, as illustrated in Figure \ref{fig:max_vol}. 
\begin{figure}[t]
	\begin{center}
		\subfloat{\includegraphics[width=3.6in]{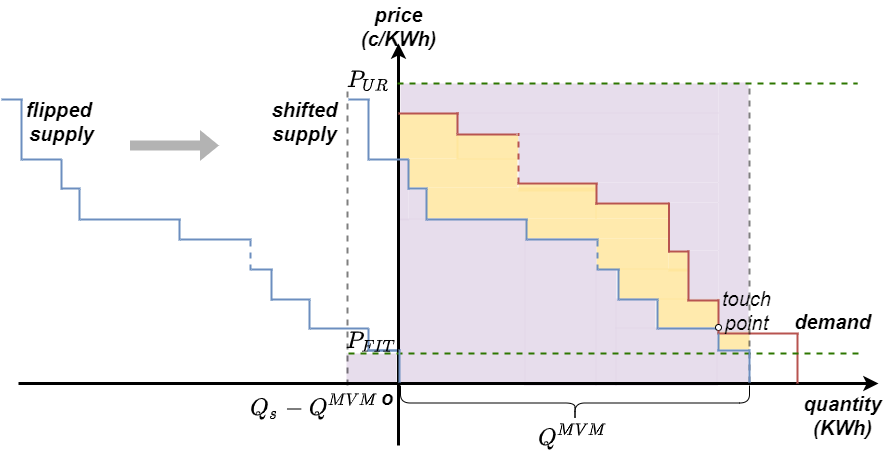}%
			\label{fig:max_vol}}\\
	\end{center}
	\caption{A maximum volume matching auction market.}
	\label{fig:max_vol}
\end{figure}
The flipped supply stack is then shifted rightward towards the demand curve until that any part of the two curves touch the first time. The distance from the origin to the right end of the flipped supply curve after shifting, denoted by $Q^{MVM}$ and illustrated in Figure \ref{fig:max_vol}, 
is exactly the maximum trading volume that the auction can achieve. 
Consequently, $Q^{MVM}$ amount of demand quantities of the highest bids are cleared; similarly, the same amount of the lowest ask quantities from suppliers are cleared. The cleared bids and asks are matched in an ascending order of prices; that is, bids and asks with the lowest prices are matched first, before moving on to the next tier of bid/ask prices. For the cleared buyer-seller pair, the buyer pays what they bid and seller gets what they ask.
Let $\mathcal{C}_b$ and $\mathcal{C}_a$  denote the set of cleared bids and asks, respectively. The uncleared supply asks (in this case, $Q_s - Q^{MVM}$) and demand bids are treated the same way as in the previous auction mechanisms. 
Hence, the total surplus of all agents is as below (represented by the light purple area in Figure \ref{fig:max_vol}):
\begin{equation}
	\begin{array}{rl}
\widehat{S}^{MVM} 
 = &  
\displaystyle \sum_{l \in \mathcal{C}_b} (P_{UR}-pb_l) \cdot qb_l 
+
\sum_{h \in \mathcal{C}_s} ps_h \cdot  qs_h \\[10pt]
& \ +\  P_{FIT}  (Q_s - Q^{MVM}).
\end{array}
\end{equation}

The auctioneer's surplus (represented by the yellow area in Figure \ref{fig:max_vol}) can be written as: 
\begin{equation}\label{eq:MVM_Auctioneer}
S^{MVM}_{au} 
= 
\sum_{l \in \mathcal{C}_b} (pb_l \cdot qb_l)
-
\sum_{h \in \mathcal{C}_a} (ps_h \cdot qs_h).
\end{equation}


\section{Analysis of the Auction Mechanisms}
\label{sec:analysis}
In this section we will analyze the theoretical properties of the four auction designs, with focuses on strategy-proofness and budget balance. While the main focus of this paper is to develop a multi-agent learning framework to implement a double-side auction, which does not rely on any of the theoretical properties, the purpose of the discussions here is to highlight the special settings of a P2P energy auction, mainly the publicly known reservation prices of both buyers ($P_{UR}$) and sellers ($P_{FIT}$), which, as we will show, can induce undesirable consequences. 

The other thing to note is that the theoretical properties discussed below are only for a single-round auction, which may not hold in a repeated setting \cite{HobbsVCG}. The task of establishing theoretical results for repeated double auction is very daunting, which again makes the multi-agent learning-based framework a valuable approach to study market outcomes in complicated settings. 

To make this paper stand-alone, we first present the definitions of the various well-established concepts in auction theory and mechanism design, which are mainly adopted from \cite{nisanalgorithmic}. To ease the presentation, we let $n$ denote the total number of buyers and sellers; that is, $n = |\mathcal{A}|$, and again drop the round superscript $h$. 

Let $A$ denote a set of alternatives. In our setting, 
$A = \{(p_i;q_i): i = 1, \ldots, n\}$, where agent $i$ can choose the price $p_i$ to bid or ask while the quantity $q_i$ is assumed to be fixed. The preference of each agent $i$ is modeled by a valuation function $v_i : A \to \mathcal{R}$, where $v_i \in V_i$, with $V_i$ being the set of possible valuation functions for player $i$. We first give the formal definition of a  mechanism.  

\newtheorem{theorem}{Theorem}[section]
\theoremstyle{definition}
\begin{definition} (Definition 9.14, \cite{nisanalgorithmic}) \label{def:mechanism}
A mechanism is a social choice function $f: V_1 \times ... \times V_n \to A$ and a vector of payment functions $p_1,...,p_n$, where $p_i: V_1 \times ... \times V_n \to \mathcal{R}$ is the amount that agent $i$ pays or receives.
\end{definition}
Given the definition of a mechanism, we are ready to introduce the definitions of strategy-proofness. 
\begin{definition} (Definition 9.15, \cite{nisanalgorithmic}) \label{def:StrategyProof}
For each $i = 1, \ldots, n$. Let $v_{-i} := (v_1,...,v_{i-1},v_{i+1},...v_n)$ be the vector with the $i-th$ component removed, and $V_{-i} := \Pi^n_{j\neq i}V_j$ be the Cartesian product of the sets $V_1, \ldots, V_n$ without $V_i$. 
A mechanism $(f,p_1,...,p_n)$ is called strategy-proof (also known as incentive compatible or truthful) if for every agent $i$, every $v'_i \in V_i$ and every $v'_{-i} \in V_{-i}$,  
$$v_i(a) - p_i(v_i , v'_{-i}) \geq v_i(a') - p_i(v'_i , v'_{-i}),$$
where $a := f(v_i,v'_{-i})$ and $a' := f(v'_i,v'_{-i})$.
\end{definition}
Note that the price $p_i(v_i,v'_{-i})$ is positive for buyers and negative for sellers. 
The above definition can be understood as follows. 
While $v_i$ represents agent $i$'s true valuation, the  agent may claim other (non-truthful) valuations $v'_i$. 
If a mechanism is strategy-proof, then the social choice $f$, 
together with the corresponding payment functions, will ensure that an agent will not be better off, in terms of the net utility $v_i(a') - p_i(v'_i , v'_{-i})$, if they do not reveal their true valuation. 

A big issue with the above definition in our specific context is that consumers (and prosumers) do not really know their valuations of electricity consumption (and generation with zero-marginal costs). Such valuations also change over time, as it is more valuable to consume energy in a hot summer day than in a calm spring night. In addition, all consumers' valuation will be higher than $P_{UR}$, as otherwise they would choose not to use electricity at all. However, if they all truthfully reveal their valuation in an auction, the auction clearing price will be higher than $P_{UR}$, which would make consumers worse-off and the auction pointless. 

While many existing works assume that consumers maximize their utility functions (valuations minus costs) to make decisions, we do not believe that such is a reasonable assumption.\footnote{We understand that the valuation also includes consumers' preference of comfortable level, such as the temperature level of an air conditioner. We will address this issue when we discuss the limitations and extensions of the current work.} The learning-based framework, as described in the previous section, does not use agents' valuation at all, which will help automate their bidding strategies. 

Without using agents' true valuation, the next best goal for an auction design is strategy-proof with respect to reservation price, as proposed in \cite{huang2002design}, which means that each agent will truthfully report their reservation price. However, while reservation prices are private information in a typical auction (such as bidding on eBay), they are clearly public information in a local P2P energy market, with $P_{UR}$ and $P_{FIT}$ being the buyers and sellers reservation prices, respectively. We will show below that in certain auction designs, such public information will result in the bang-bang type of market clearing prices; that is, the clearing prices will be either $P_{UR}$ or $P_{FIT}$, depending on the total supply and demand quantities.


To further compare the outcomes of the four auction designs, we introduce the additional  concepts of budget balance, which is about if the auctioneer gets positive payments in an auction.
\begin{definition} (\cite{nath2019efficiency,huang2002design})
A mechanism is budget-balanced if it gets zero payment; that is, $\sum_{i=1}^n p_i(v_1, ..., v_n) = 0$ for every $v_i \in V_i$, $i= 1, \ldots, n$. We call it weakly budget-balanced if it gets non-negative payment from agents; that is,  $\sum_{i=1}^n p_i(v_1, ..., v_n) \geq 0$ for every $v_i \in V_i$, $i= 1, \ldots, n$.
\end{definition}

In the following, we will analyze the above-defined concepts for each of the four auction mechanisms introduced in the previous section.

\subsection{$k$-Double Auction}\label{subsec:kDoubleProp}
We first show that the mechanism of $k$-double auction is not strategy-proof with respect to reserve prices.  For the following discussions, any agent $i$ (buyer or seller)'s utility is defined as $\Lambda_i - \underline{\Lambda}_i$, with $\Lambda_i$ and $\underline{\Lambda}_i$ defined in Equation \eqref{eq:rwd_origin} and \eqref{eq:rwd_low}, respectively. In addition, we lay out the following assumptions.  

\begin{assumption}\label{as:TrueQ}
All buyers and sellers truthfully submit their quantities to buy or sell.
\end{assumption}

\begin{assumption}\label{as:KnownReserve}
The reservation prices of buyers and sellers $P_{FIT}$ and $P_{UR}$, respectively, are both public information. 
\end{assumption}

\begin{proposition}\label{prop:kNotIC}
Consider a single round $k$-double auction with $k \in [0, 1]$, where buyers and sellers only bid prices. Assume that Assumption \ref{as:TrueQ} and \ref{as:KnownReserve} hold. 
Furthermore, let $\widehat{Q}_s$ and $\widehat{Q}_b$ denote the total supply and demand quantities (not just cleared bids) in the auction, respectively. If the relationship between $\widehat{Q}_s$ and $\widehat{Q}_b$ is also public information (that is, $\widehat{Q}_s \geq \widehat{Q}_b$ or $\widehat{Q}_s\leq \widehat{Q}_b$), then the $k$-double auction is not strategy-proof with respect to reservation prices. 
\end{proposition}
\begin{proof} If all agent bids truthfully with respect to the reservation price, then the market clearing price in the $k$-double auction will be $kP_{UR} + (1 - k)P_{FIT}$. For the case where $\widehat{Q}_s \leq \widehat{Q}_b$, consider a seller $g$: if all other sellers ask $P_{FIT}$, and all buyers bid $P_{UR}$, then seller $g$ can ask $P_g$, with $P_{FIT}< P_g < P_{UR}$ for all his/her sell quantities. All the quantities will still be cleared, since $\widehat{Q}_s \leq \widehat{Q}_b$. Then the market clearing price would be $kP_{UR} + (1 - k)P_h$, strictly greater than $kP_{UR} + (1 - k)P_{FIT}$, and hence, seller $g$'s utility is strictly higher. Similar arguments can be made for buyers when $\widehat{Q}_s \geq \widehat{Q}_b$. 
\end{proof}
When $k$ = 0 or 1, Assumption 2 actually can be dropped, meaning that even without knowing the actual supply and demand situation, the agents still do not have incentives to bid the reserve price. The proof is a simple extension of the above argument and is omitted here.

The non-strategy-proofness of the $k$-double auction is not surprising, as in such a mechanism, the marginal bidders (namely, the lowest bids or highest sellers who are cleared in an auction) set the market clearing prices, which gives agents incentives to manipulate the clearing prices in their favor. 
The Vikerey-like auctions as introduced in the previous section are exactly designed to overcome such an issue. 


While Proposition \ref{prop:kNotIC} is about the property of the auction mechanism, in the following we study from the perspective of the agents, and discuss what bidding strategies can lead to an ex-post Nash equilibrium, as  defined below.  

\begin{definition} (Definition 9.22, \cite{nisanalgorithmic}) \label{def:ExPostNash}
	A profile of strategies $s_1, \ldots, s_n$ of $n$ agents is an ex-post Nash equilibrium if for all $i=1, \ldots, n$, all types $y_i \in Y_i$, and all feasible actions $a'_i$ of $i$, we have that 
	$\pi_i(y_i, s_i(y_i), s_{-i}(y_{-i})) \geq  
	\pi_i(y_i, a'_i, s_{-i}(y_{-i}))$, where $\pi_i$ is $i$'s utility function. 	
\end{definition}
In essence, an ex-post Nash equilibrium requires that $s_i(y_i)$ is the best response to  $s_{-i}(y_{-i})$ for all possible $y_{-i}$, which is the collection of other agents' types.	

\begin{proposition}\label{prop:ExPostNash}
Under the same context and assumptions in Proposition  \ref{prop:kNotIC}, with a given $k\in [0,1]$, we have the following results.
\begin{itemize}
\item[(1)] If $\widehat{Q}_s > \widehat{Q}_b$, all agents bidding/asking $P_{FIT}$ is an ex-post Nash equilibrium. 

\item[(2)] If $\widehat{Q}_s < \widehat{Q}_b$, all agents bidding/asking $P_{UR}$ is an ex-post Nash equilibrium. 

\item[(3)] If $\widehat{Q}_s = \widehat{Q}_b$, and let $\widetilde{P}$ be a given price in $[P_{FIT}, P_{UR}]$. 
Then all agents bidding/asking $\widetilde{P}$ is an ex-post Nash equilibrium. 
\end{itemize}
\end{proposition}
\begin{proof}
The proofs for all the three situations are similar; hence, we only show the proof of case (3) here. In (3), the strategy for any agent (buyer or seller) is that $s_i(y_i) = \widetilde{P}$, for all $i\in \mathcal{A}$, and all $y_i \in Y_i$. With this strategy, all buyers' and sellers' quantities are cleared (since $\widehat{Q}_s = \widehat{Q}_b$). 
Now assume that a particular buyer $j$ chooses $pb_j \neq \widetilde{P}$. If $pb_j > \widetilde{P}$, buyer $j$'s bid will be cleared, but the market clearing price is still $\widetilde{P}$, and buyer $j$'s utility is the same as bidding $\widetilde{P}$. If $pb_j < \widetilde{P}$, 
$j$'s bid will not be cleared, since all sellers ask $\widetilde{P}$. Then buyer $j$'s utility is zero, strictly less than bidding $\widetilde{P}$. The arguments for the seller side are the same. Hence, the strategy $s_i(y_i) = \widetilde{P}$, for all $i\in \mathcal{A}$, and all $y_i \in Y_i$ is an ex-post Nash equilibrium.
\end{proof} 
Note that under situation (1) and (2) in Proposition \ref{prop:ExPostNash}, the market outcomes are susceptible to the bang-bang type results; that is, either the buyers or the sellers will reap all the benefits, depending on the total supply versus total demand, leaving the other parties of zero benefits (compared to directly buying from or selling to a utility company). Under situation (3), the result in Proposition \ref{prop:ExPostNash} does not appear to be useful, as $\widetilde{P}$ can be anything between $P_{FIT}$ and $P_{UR}$, and there is no way for the agents to agree upon a common point a priori. However, the public information of the two reserve prices for buyers and sellers, respectively, provide a focal point in a $k$-double auction; that is, $\widetilde{P} = (P_{FIT} + P_{UR})/2$. This is indeed what we observe in our simulations, along with the outcomes as predicted by Proposition \ref{prop:ExPostNash} under situation (1) and (2). 


Regarding the other property, the $k$-double auction is strongly budget-balanced because the selling and buying quantity are equal, and the clearing price is the same for buyers and sellers. This means that the auctioneer's surplus is always zero in a $k$-double auction, which can be a desirable outcome in some cases, but can be a downside here. As the role of an auctioneer in a P2P energy market is likely played by a utility company or a DSO, they may require a payment to provide such service. A double auction may run on a distributed ledger system (aka Blockchain) without a central auctioneer. Even then, distribution of the cleared energy from sellers to buyers need access to the physical distribution networks, which entail maintenance and repair costs. If such costs need to be recouped from the auction process, then a weakly budget-balanced mechanism, such as the Vickrey variant auction, can be a choice. 

\subsection{Vickrey Variant Double Auction}
\label{subsec:VVProp}

The key difference between a $k$-double auction and a Vickrey-like auction is that the marginal winners in the later do not affect clearing prices. Hence, the strategy-proofness of a Vickrey auction still holds in a multi-unit, double auction setting, as proved in \cite{huang2002design}, along with the fact that it is weakly budget-balanced.

While the truthfulness of an auction is usually a desired property, it is actually the opposite in a P2P energy market. This so because the goal of a local energy trading market is different than a traditional auction. In a traditional auction, the objective is to allocate resources to people who value them the most (and a private reservation price of each bidder can be used as a proxy for their valuation of the resources). In a local energy market, everyone needs electricity, and the goal is not about efficient allocation. Instead, the P2P market is to help buyers and sellers achieve more favorable rates than $P_{UR}$ and $P_{FIT}$, respectively. 

While the theorem predicts that in our specific setting, the outcomes of the Vickrey variant double auction will be always be $P_{UR}$ for buyers and $P_{FIT}$ for sellers, we will see that this is not always the case in our numerical results. This seeming contradiction highlights the point that all the theoretical results in this section are only established for a one-shot auction, while they may not be true in a repeated game. In fact, it is known that a Vickrey auction is susceptible to tacit collusion in a repeated auction \cite{HobbsVCG}, and the truthfulness property no longer holds.

\subsection{McAfee Double Auction}
\label{subsec:McAfeeProp}
The strategy-proofness of the McAfee auction for a single-unit good is provided in \cite{mcafee1992dominant}. Here we extend the proof to a double auction with multi-unit divisible goods.
\begin{proposition}
Under Assumption \ref{as:TrueQ}, the McAfee double auction is strategy-proof with respect to reservation prices. In addition, it is weakly budget-balanced.
\end{proposition}
\begin{proof}
	As in the description of the Vickrey-variant double auction, there are two cases regarding to the ordering of the supply and demand bid curves and the corresponding quantities. They are referred to as Case I and Case II, and are represented by Equation \eqref{eq:CaseI} and \eqref{eq:CaseII}, respectively. It suffices to show the proof for Case I, as the arguments for Case II will be the same. Under Case I, there are still two cases: Case A and Case B, as in the description of the McAfee auction. In either case, there is no incentive for buyers to bid above $P_{UR}$. Hence, for $pb_L$, the last price bid before the supply and demand curve intersects, we have  $ pb_L \leq P_{UR}$. For an arbitrary buyer $i$, assume that the buyer do not truthfully bid $P_{UR}$; that is, $pb_i< P_{UR}$. If $pb_i > pb_L$, buyer $i$ will have the same utility as biding $P_{UR}$, since this underbidding will not affect  the market clearing price or quantity. If $pb_i = pb_L$, under Case A, buyer $i$ will be cleared and the market clearing price is still not affected, since the clearing price is determined by $p_{L+1}$ and $P_{H+1}$. Under Case B, buyer $i$ will not be cleared, and hence, is strictly worse off than bidding $P_{UR}$. If $pb_i \leq pb_{L+1}$, buyer $i$ will not be cleared in either case, and hence worse off than bidding $P_{UR}$. The arguments for sellers are similar. Hence, by Definition \ref{def:StrategyProof}, we have shown that the McAfee auction is strategy-proof with respect to reservation prices.

Under Case A, the McAfee auction will result in the same clearing price for buyers and sellers, the same as in a $k$-double auction. Hence, it is strongly budget-balanced; while under Case B, it is the same as the Vickrey-variant auction, which is weakly budget-balanced as proven in \cite{huang2002design}. Hence, overall the McAfee auction is weakly budget-balanced.
\end{proof}

\subsection{Maximum Volume Matching Double Auction}
\label{subsec:MVMProp}

The mechanism of MVM double auction is not strategy-proof (with respect to reservation price). By Definition \ref{def:StrategyProof}, it suffices to find one particular instance under which an agent has an incentive to be not truthful. One such instance is that  when all other buyers bid $P_{UR}$ except buyer $i$, and all sellers ask $P_{FIT}$. Suppose that the total supply is greater than total demand. If buyer $i$ bids any price strictly between $P_{UR}$ and $P_{FIT}$, the quantity allocated to the agent remains the same and the price buyer $i$ pays is lower than $P_{UR}$, since the MVM is pay-as-bid. As a result, buyer $i$ will get strictly better utility of not bidding  $P_{UR}$; hence, the MVM double auction is not strategy-proof with respect to reservation price. 

On the other hand, since the buying price is always no lower than the selling price for each matching (as shown in Fig. \ref{fig:max_vol}), 
the auctioneer's payoff, as given in \eqref{eq:MVM_Auctioneer}, is always non-negative, which means that the MVM double auction is weakly budget-balanced. 
\section{Numerical Simulations}
\label{sec:num_res}
In this section, we apply the MAB game approach to compare the four different auction mechanisms in an energy market. 

\subsection{Input Data}
\label{sec:sim_data}

\subsubsection{Decision epochs and temporal resolution}

As a starting point, we consider daily auctions with an hourly temporal resolution. More specifically, one round of an auction happens in a specific hour, such as 9 AM - 10 AM, which is repeated daily. Other hours, such as 10 AM - 11 AM, are considered as as different auctions, and agents will learn different strategies. In our simulations, we perform seven auctions in a day, representing the seven hours between 9 AM and 4 PM, and we run for 365 days. 
\subsubsection{$P_{UR}$, $P_{FIT}$ and the decision space}
$P_{UR}$ and feed-in tariff $P_{FIT}$ are fixed throughout the simulation, and are set at $P_{UR} = 11$  $\cent$/KWh and $P_{FIT} = 5$ $\cent$/KWh. 
For price bids/asks, to account for bounded rationality from both buyers and sellers, we disretize the price range from $0$ $\cent$/KWh to $14 $ $\cent$/KWh, with 1 $\cent$/KWh increment. Doing so is to allow the sellers to ask bellow $P_{FIT}$ or for buyers to bid above $P_{UR}$. While such behavior is not rational,   if an agent believes that the market clearing price will be between $P_{FIT}$ and $P_{UR}$, then asking below $P_{FIT}$ for sellers and bidding above $P_{UR}$ for buyers will almost certainly guarantee the clearing of their bids. However, if all agents do so, the market clearing price can fall outside of $[P_{FIT}, P_{UR}]$, which we indeed observed in certain rounds in our simulation. The corresponding reward to the agent will be 0, per the definition of \eqref{eq:rwd_function}. The learning algorithm will then make such strategies unlikely (but not entirely impossible) to be chosen again. The modeling flexibility of the learning-based approach allows us to study a wide range of bidding strategies/behavior without having to make any unrealistic assumptions on agents' beliefs, knowledge, or rationality.

\subsubsection{Energy consumption, supply and types of agents} 
As described in Section \ref{subsec:AgentType}, we simulate three groups of agents: 1,000 pure consumers, 1,000 pure suppliers, and 500 prosumers. For the last group, the net position of their supply and demand in a particular hour determines their roles in the corresponding round of the auction. For DER generation, we assume that is all from rooftop photovoltaic (PV) panels. The energy demand and generation of each buyer and seller are all randomly generated based on the System Advisor Model (SAM) \cite{nrelsam} National Renewable Energy Laboratory (NREL). The details of how such data are generated, along with the data we used in our simulation, are available on Github.  \footnote{\url{https://github.com/feng219/MAB_Algorithm}} 
In addition, at each round, each agent has a 0.005 of being regenerated independently. The learning history of the agent will be cleared to all-zero vector after regeneration.

\subsection{Individual Agent's Bandit Learning Algorithm}
In our simulation, each agent has an equal probability of using one of the three algorithms to choose a specific arm in each round: UCB1, UCB2, and $\epsilon$-greedy. Once an algorithm is chosen, the agent will use the same algorithm throughout the repeated auction till regeneration. A regenerated agent again will have equal probability to choose from one of the three algorithms. The UCB1 and UCB2 algorithms are developed in \cite{auer2002finite}, and the detailed codes in our simulation are available on the same Github location.

\subsection{Numerical Results}
This subsection reports the simulation results of the four different auction designs. Note that for the $k$-double auction, we only consider the case where $k=0.5$, and refer it as $k05$ double auction in this subsection.

\begin{figure}[!h]
	\centering
		\centering
		\includegraphics[width=6.5cm]{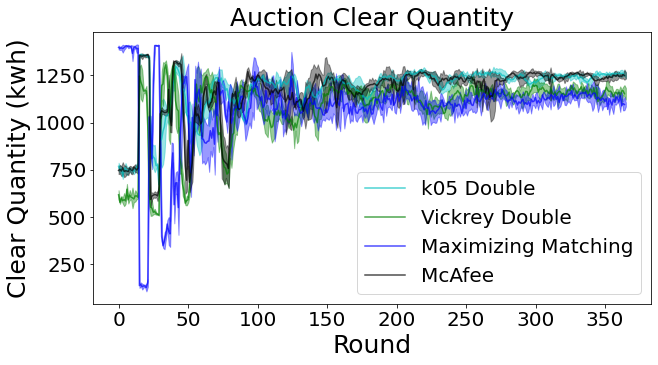}
		\caption{Total clearing quantity (KWh) in the 9 AM - 10 AM auction.}
		\label{fig:clearing_quant}
\end{figure}

\begin{figure}[!h]
	\centering
	\includegraphics[width=6.5cm]{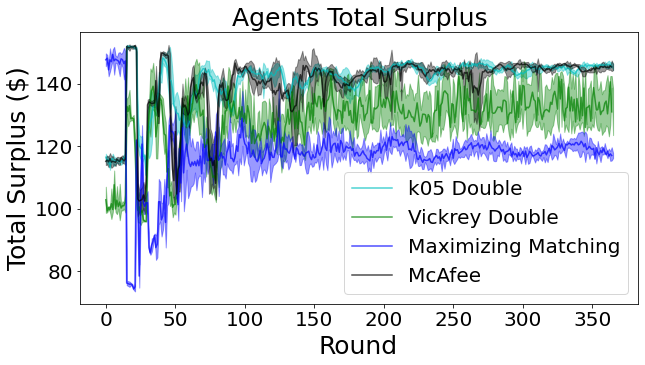}
	\caption{The sum of buyers and sellers' surplus (\$) in the 9 AM - 10 AM auction.}
	\label{fig:Social_welfare}
\end{figure}

\begin{figure}[!h]
\centering
	\centering
	\includegraphics[width=6.5cm]{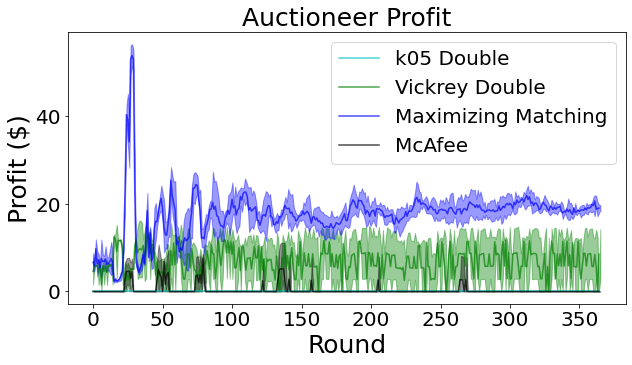}
	\caption{Auctioneer's profit (\$) in the auctions (9 am)}
	\label{fig:Auctioneer_profit}
\end{figure}

In the simulation, for each auction (such that the auction for 9 - 10 AM on each day), we run four times with different random seeds, with each time consisting of 365 rounds of the auction. 
The codes are written in Python and run on a PC with Windows 10 OS, Intel Core i7-6700K processor, and 16 GB RAM. The time of one particular run (i.e., one time of the 365 rounds) of the $k$-double auction, Vickrey-variant, McAfee and MVM auction are 281, 285, 293, and 276, respectively. The time for the other three runs are similar. We first show results corresponding to one particular hour, 9 - 10 AM, since the observed trends of market outcomes are exactly the same in the other hours. 
The solid lines in Figures \ref{fig:clearing_quant}, \ref{fig:Social_welfare}, and \ref{fig:Auctioneer_profit}  represent the average of the four simulation runs; while the shades in the figures represent the values within one standard deviation of the mean. 

Fig. \ref{fig:clearing_quant} compares the cleared quantities from the four different auction mechanisms. It can be seen that the Vickrey-variant (VV) auction results in lower cleared quantities than $k05$-double and McAfee. This is expected as the VV mechanism design sacrifices some traded volume (as it only clears up to the $L$-1th buyer and $H$-1th seller) to achieve strategy-proofness. What is surprising is that the cleared quantity of the MVM auction, which is designed (and proved in \cite{niu2013maximizing}) to maximize the traded volume, is also lower than $k05$-double and McAfee. The reason is as follows: in a single-round auction, with everything being equal, the MVM auction will be able to reach the theoretical upper bound of the maximum cleared quantity. However, in a dynamic setting with the learning-based approach, agents will learn different things in different auction designs. Hence, at the beginning of each round (especially in the later rounds), the assumption that ``everything being equal" no longer holds, which results in the observed outcomes that on average, the cleared quantity in MVM auction can be less than other auction designs. The cleared quantities between $k05$ double and McAfee auctions are very similar. 

Next we compare the total surplus, i.e., the $\widehat{S}$ as defined in Section \ref{sec:double_auction}, in Fig.
 \ref{fig:Social_welfare}. 
It can be seen in Fig. \ref{fig:Social_welfare} that total surplus of the VV auction is notably less than the $k05$ double auction, exactly for the same reason as why the cleared quantity in the VV auction is less. Since the McAfee auction is a hybrid of the $k$-double and the VV auction, agents in the McAfee auction can likely learn the surplus differences between the $k$-double and the VV auction through repeated interactions, and will make the McAfee auction outcome the same as the $k$-double auction most of the time.  The total surplus is the lowest in the MVM auction, which likely is due to the surplus transfer from agents to the auctioneer, as shown in Figure \ref{fig:Auctioneer_profit}. 

Fig. \ref{fig:Auctioneer_profit} shows the auctioneer's surplus from the four auction mechanisms. Clearly the MVM auction results in the most auctioneer surplus, which helps explain why the agents' surplus in the MVM auction is the lowest. 
 The VV auction also results in positive auctioneer's payoff as buyers pay more than what sellers receive. The $k$-double auction, regardless $k$'s value always results in zero auctioneer surplus. For McAfee auction, again, due to its hybrid nature, the auctioneer does have positive surplus over the course of the repeated auction, but the surplus is significantly less than the VV auction. 

\begin{figure}[!h]
	\centering
	\includegraphics[width=6.5cm]{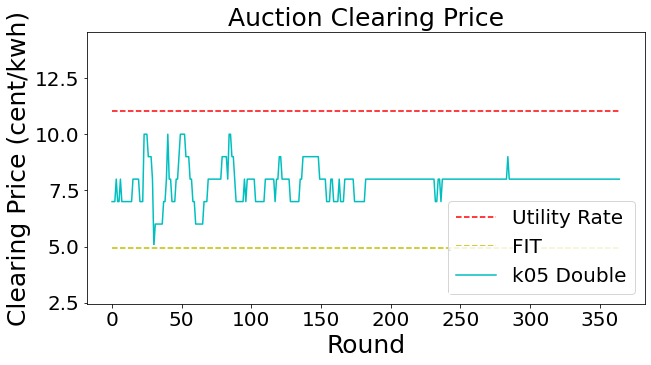}
	\caption{k05 auction clearing price (9 AM) over the rounds.}
	\label{fig:k05 Price 9 AM over rounds}
\end{figure}

Based on the numerical results, it appears that the $k$-double auction and the McAfee auction are the better performing mechanisms, so long as that an auctioneer's payoff is not a concern. To have a more complete picture, we present the clearing prices from the $k05$ auction in chronological order over seven days. All the results shown below are taken from the later rounds of the simulation when the market outcomes appear to stabilize. Note that the clearing prices are taken from just one of the four simulation runs, not the average of the four runs, as all the runs exhibit the same patterns. 
\begin{figure}[!h]
\centering
	\centering
	\includegraphics[width=6.5cm]{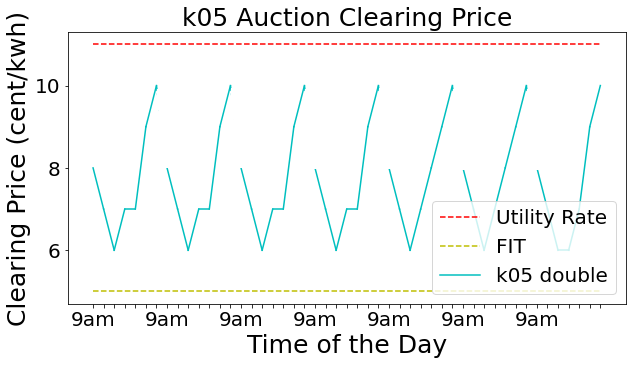}
	\caption{k05 auction clearing price (consecutive 7 days)}
	\label{fig:k05 Auction clearing Price (buy)_ts}
\end{figure}
The pattern of the auction clearing price closely resembles that of the demand/supply ratio in the market, as shown in Fig. \ref{fig:Demandsupply}.
\begin{figure}
	\begin{center}
		\subfloat{\includegraphics[scale = 0.35]{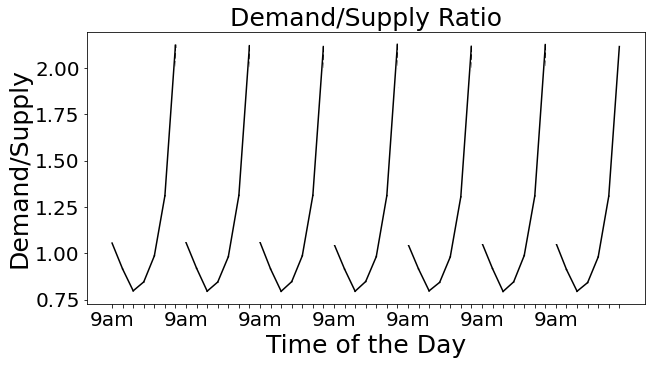}%
			\label{fig:Demandsupply}}\\
	\end{center}
	\caption{Demand/supply ratio (consecutive 7 days).}
	\label{fig:Demandsupply}
\end{figure}
In our simulations, the demand/supply ratio is close to 1 at 9 AM, as shown in Fig. \ref{fig:Demandsupply}. When it is around noon time, however, the supply is more than the demand (due to high PV generation and low demand at residential home, as people are out at work), and the corresponding clearing price in the $k05$ auction gets close to $P_{FIT}$. In late afternoons, as the PV production winds down and residential demand picks up, the clearing price approaches to $P_{UR}$. These two situations are exactly as what predicted in Proposition \ref{prop:ExPostNash}.  
To have a closer look of the market outcomes when the demand/supply ratio is 1, we present the clearing prices for the 9 - 10 AM auction of the $k05$ auction over the entire simulated horizon in Fig. \ref{fig:k05 Price 9 AM over rounds}. 
As seen in Fig. \ref{fig:k05 Price 9 AM over rounds}, 
the undesirable bang-bang type of clearing prices (that is, either $P_{UR}$ or $P_{FIT}$) did appear at the initial rounds of the auction. As the learning progresses, the market clearing prices tend to stabilize at around 8 $\cent$/kwh prior to 200 rounds, 
which is exactly half way between $P_{UR}$ (11 $\cent$/kwh) and $P_{FIT}$ (5 $\cent$/kwh). This is consistent to what we speculated when discussing the results of Proposition \ref{prop:ExPostNash}. 
 We consider this market outcome fair as it splits the total surplus 
 equally between buyers and sellers. 
\begin{figure}[!h]
\centering
	\centering
	\includegraphics[width=6.5cm]{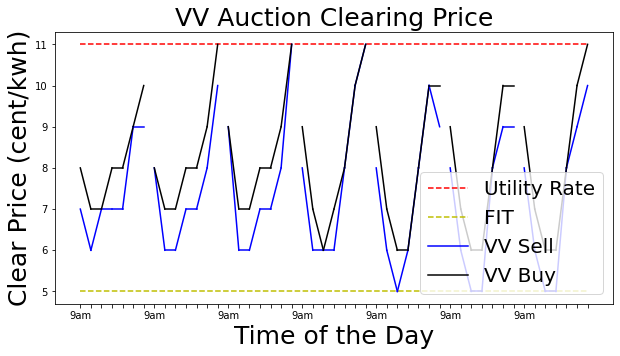}
	\caption{Vickrey-variant auction clearing price over 7 days}
	\label{fig:VVprice}
\end{figure}

\begin{figure}[!h]
\centering
	\centering
	\includegraphics[width=6.5cm]{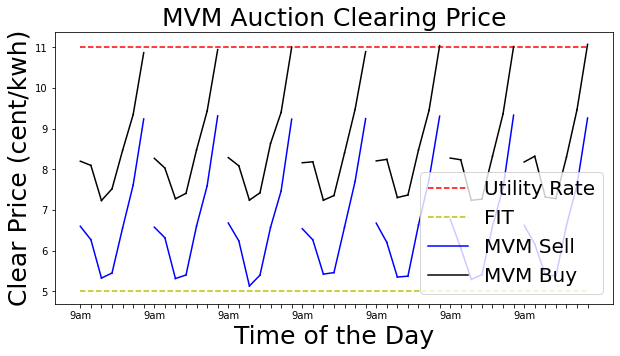}
	\caption{MVM auction clearing price over 7 days}
	\label{fig:MVMprice}
\end{figure}
The outcomes of the McAfee double auction are very similar to the $k05$ auction. The clear prices of the Vickrey-variant and maximum volume matching auctions are shown in Fig. \ref{fig:VVprice} and \ref{fig:MVMprice}, respectively.\footnote{Note that for the MVM auction, since it is pay-as-bid, the prices reported are volume-averaged; that is, for all cleared buyers, the average clearing price is $\sum_{l\in C_b} (pb_l * qb_l)/\sum_{l\in C_b} qb_l$, and the same for the cleared sellers.} 
The results of the Vickrey-variant double auction exhibit similar trends following the demand/supply ratio in a day, but the clearing prices are more extreme than the $k05$ auction when demand and supply are imbalanced. This is so because the averaging mechanism in determining the clearing price in a $k05$ auction helps alleviate the extreme prices.


\section{Conclusion and Future Research}
\label{sec:conclusion}
In this work, we proposed a framework based on multi-agent MAB-games to help automate the bidding strategies for consumers and prosumers participate a P2P energy market. This approach also provides a framework to study complicated games, such as repeated games with incomplete information, where theoretical results are either scarce or of little practical usage. 
The approach has shown to be very useful in numerically comparing market designs, as we applied it to compare four specific implementations of a double auction in an energy trading market, which can be a handy tool for policy makers to test their market design before actual implementation. The framework is also very flexible as it can easily incorporate different kinds of heterogeneity of agents, and it requires minimal storage and computing power on the agent side, which should help the employment of such an approach among real-world consumers/prosumers. 

Independent of the MAB-game framework, we studied the theoretical properties of four specific double-auction designs in the context of a local P2P energy market, which presents the distinct features of all zero-marginal-cost supplies and publicly known reserve prices for both buyers and sellers. We showed that such features can be undesirable as they may lead to bang-bang type auction clearing prices, depending on the demand/supply ratio in the market and auction  mechanism. These results also highlight the needs of sophisticated simulation framework that can capture the essence of repeated double auctions with a large number of agents with bounded rationality (such as due to lack of expertise knowledge, information, or computing resources), as the theories need to be tested to see if they can indeed emerge from repeated interactions among agents. 

The current work only means to be a starting point for the general topic of decentralized multi-agent games and their applications in better utilizing DERs energy systems. It has several notable limitations and can certainly benefit from future research. First and foremost, the presented work does not consider physical network constraints, which are no double essential for any P2P market design to be implementable in the real-world. We point out this is not necessarily a limitation of the MAB-game framework. The auction mechanism can be used to pre-commit resources some time ahead, say, an hour ahead, exactly the same way as the day-ahead market in the wholesale market. After each clearing, a utility company or a distributed system operator (DSO) can run distribution-level power flow equations to determine if the cleared bids are physically feasible. If not, the past round of auction can be re-run, and each cleared agent will update their rewards of the last round to zero so that the choice is likely to be picked in the future. The MAB-game approach is used to help consumers/prosumers respond to real-time pricing in \cite{zhao2018electricity}, which includes a system operator's optimal power flow problem to determine locational marginal prices. The results there show show that agents can indeed learn to alleviate transmission congestion to earn better payoffs. Granted, such an approach still cannot address real-time feasibility issues, which are likely to be dealt with in a completely different framework. In addition, specific rules or mechanisms need to be designed to compensate power losses and distribution network maintenance costs. 

Another notable limitation of the MAB-game approach is that it cannot easily handle time-linkage decisions, such as injection/withdraw decisions for energy storage resources. We are currently developing a multi-agent reinforcement learning framework to include energy storage in repeated auctions. Along this direction, we can also consider thermal load (HVAC) and consumers' preference of comfort (such as reflected by temperature settings). Such works will be reported in follow-up papers.


\bibliographystyle{ieeetr}
\bibliography{MAB_Auction}

\end{document}